\documentclass[letter]{aa}

\usepackage{graphicx,xspace}
\usepackage{color}
\usepackage{graphicx,color}
\usepackage{natbib}
\usepackage{microtype}
\usepackage{url}
\usepackage[varg]{txfonts}
\usepackage{threeparttable}
\usepackage{booktabs, makecell}
\usepackage{longtable}
\usepackage[utf8]{inputenc}
\usepackage{amsmath}
\usepackage{verbatim}
\usepackage{txfonts}

\def\ero{\textit{eROSITA}\xspace}

\def\gai{\textit{Gaia}\xspace}

\def\ros{\textit{ROSAT}\xspace}
\def\srgero{\textit{SRG}/eROSITA\xspace}

\def\swi{\textit{Swift}/XRT\xspace}
\def\tes{\textit{TESS}\xspace}
\def\xmmn{\textit{XMM-Newton}\xspace}
\def\xspec{\texttt{\textit{XSPEC}}\xspace}

\newcommand\targ{{J1912-44}\xspace}
\newcommand\erj{{eRASSU~J191213.9$-$441044}\xspace}
\newcommand\arsc{{AR~Sco}\xspace}

\newcommand\fergs{\ensuremath{\mathrm{erg}\,\mathrm{cm}^{-2}\,\mathrm{s}^{-1}}\xspace}
\newcommand\lum{\ensuremath{\mathrm{erg}\,\mathrm{s}^{-1}}\xspace}
\newcommand\fcgs{\ensuremath{\mathrm{erg}\,\mathrm{cm}^{-2}\,\mathrm{s}^{-1}\,\mbox{\AA}^{-1}}\xspace}
\newcommand\ecf{\ensuremath{\mathrm{erg}\,\mathrm{cm}^{-2}\,\mathrm{\AA}^{-1}\,\mathrm{ct}^{-1}}\xspace}

\newcommand\rat{\ensuremath{\mathrm{s}^{-1}}\xspace}

\begin{document}

\title{X-ray properties of the white dwarf pulsar \erj\thanks{Based on observations obtained with \xmmn, an ESA science mission with instruments and contributions directly funded by ESA Member States and NASA}}

\author{Axel Schwope\inst{1}
\and
T.R.Marsh\inst{2}
\and
Annie Standke\inst{1,3}
\and
Ingrid Pelisoli\inst{2}
\and
Stephen Potter\inst{4,5}
\and
David Buckley\inst{4,6,7}
\and
James Munday\inst{2,8}
\and
Vik Dhillon\inst{9,10}
}
\institute{Leibniz-Institut f\"ur Astrophysik Potsdam (AIP), An der Sternwarte 16, 14482 Potsdam, Germany\\
\email{aschwope@aip.de}
\and
Department of Physics, University of Warwick, Gibbet Hill Road, Coventry, CV4 7AL, UK
\and
Potsdam University, Institute for Physics and Astronomy, Karl-Liebknecht-Stra\ss e 24/25, 14476 Potsdam, Germany
\and
South African Astronomical Observatory, PO Box 9, Observatory Road, Observatory 7935, Cape Town, 
\and
Department of Physics, University of Johannesburg, PO Box 524, 2006 Auckland Park, Johannesburg, South Africa
\and
Department of Astronomy, University of Cape Town, Private Bag X3, Rondebosch 7701, South Africa
\and
Department of Physics, University of the Free State, PO Box 339, Bloemfontein 9300, South Africa
\and
Isaac Newton Group of Telescopes, Apartado de Correos 368, E-38700 Santa Cruz de La Palma, Spain
\and
Department of Physics and Astronomy, University of Sheffield, Sheffield S3 7RH, UK
\and
Instituto de Astrof\'{i}ısica de Canarias, E-38205 La Laguna,
Tenerife, Spain}

\titlerunning{X-ray properties of the white-dwarf pulsar \erj}
\authorrunning{Schwope et al. }

\date{\today}

\keywords{stars: binaries: close; stars: individual: \erj; X-rays: binaries}

\abstract{We report X-ray observations of the newly discovered pulsating white dwarf \erj with Spectrum Roentgen Gamma and eROSITA (\srgero)  and \xmmn. The new source was discovered during the first \ero all-sky survey at a flux level of $f_{\rm X} (0.2 - 2.3 \,\mbox{keV}) = 3.3\times 10^{-13}$\,\fergs and found to be spatially coincident with a $G=17.1$ stellar \gai-source at a distance of 237\,pc. The flux dropped to about $f_{\rm X} = 1 \times 10^{-13}$\,\fergs during the three following \ero all-sky surveys and remained at this lower level during dedicated \xmmn observations performed in September 2022. With \xmmn, pulsations with a period of 319 s were found at X-ray and ultraviolet wavelengths occurring simultaneously in time, thus confirming the nature of \erj as the second white-dwarf pulsar. 
The X-ray and UV-pulses correspond to broad optical pulses. Narrow optical pulses that occurred occasionally during simultaneous XMM-Newton/ULTRACAM observations have no X-ray counterpart. The orbital variability of the X-ray signal with a roughly sinusoidal shape was observed with a pulsed fraction of $\sim$28\% and maximum emission at orbital phase $\sim$0.25. The ultraviolet light curve peaks at around binary phase 0.45. The X-ray spectrum can be described with the sum of a power law spectrum and a thermal component with a mean X-ray luminosity of $L_{x} \mbox{(0.2-10 keV)} = 1.4 \times 10^{30}$\,erg\,s$^{-1}$. The spectral and variability properties could indicate some residual accretion, in contrast to the case of the prototypical object AR~Sco.
}

\maketitle

\section{Introduction}
\arsc, identified in 2016 by \cite{marsh+16}, is a highly peculiar white dwarf main sequence (WDMS) binary, composed of a strongly magnetic white dwarf and an M-dwarf companion orbiting each other with a period of 3.56 hours \citep{buckley+17}. The spin of the WD is only 1.95 minutes and pulsed emission on the 1.97-min beat period has been observed from radio to X-ray wavelengths. \arsc is thought to be powered by the spin-down of the white dwarf, and given the observed pulsed and highly polarimetric signal is said to be of synchrotron origin, the object has been deemed as the first white dwarf pulsar. Its X-ray properties were analysed by \cite{takata+18, takata+21}, who found a pronounced variability of the X-ray brightness for both the orbital and the beat period between spin and orbital periods. The orbital-phase dependent maximum was observed at superior conjunction of the M star, an observation used to locate the physical origin of the emission on the irradiated companion star. 

\begin{figure*}[t]
\resizebox{0.47\hsize}{!}{\includegraphics{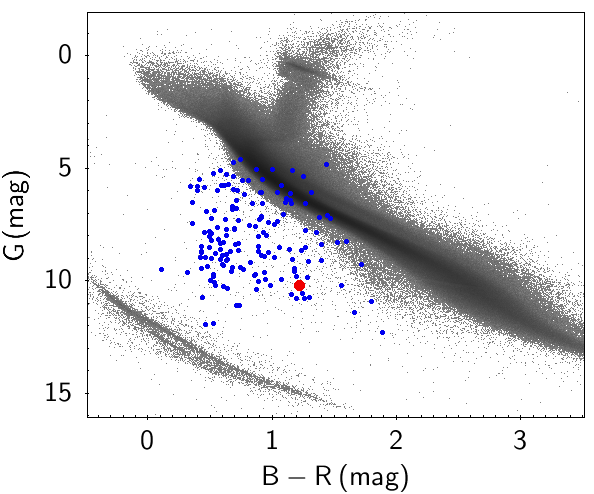}}
\hfill
\resizebox{0.49\hsize}{!}{\includegraphics{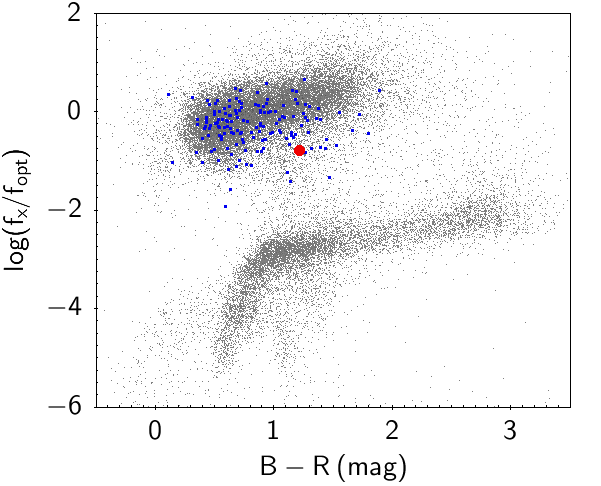}}
\caption{Color-magnitude diagram of 161 CB candidates drawn from matching eRASS1 -- \gai sources (blue symbols) shown on the left.  The background is made of $\sim$230,000 unrelated \gai sources with $\pi/\sigma_\pi > 100$ ($\pi$: parallax). X-ray to optical flux ratio ($\log{f_{\rm X}} + \mbox{phot\_g\_mean\_mag}/2.5 + 4.86$) vs.~optical colour of eRASS1 sources matching with \gai within a radius of 1 arcsec, shown on the right. The lower branch on the right figure is populated by coronal emitters, the upper branch by AGN and accreting binaries. The position of \targ is marked red in both diagrams and the 161 \gai candidates in blue.
\label{f:diag}}
\end{figure*}

\arsc (and \arsc-like objects) seem to play an important role in the evolution of close binaries and  thereby contribute greatly to our understanding of how strong magnetic fields are generated in white dwarfs. According to \cite{schreiber+21}, the field is generated when the WD crystallizes, transforming a previously semi-detached non-magnetic cataclysmic binary (for a certain time interval) into a detached WDMS binary. With just one object of this kind known to date, it is difficult to find clear observational support for this scenario. The recent discovery of an \arsc-twin, J191213.72$-$441045.1 (henceforth, \targ) by \cite{pelisoli+23}, with $P_{\rm orb} = 4.03$\,h and $P_{\rm spin} = 5.32$\,min has established \arsc-like objects as a class, lending support to this evolutionary picture. \cite{pelisoli+23} found pulsed emission from radio to X-ray wavelengths and pronounced polarimetric variability on the spin period. The spin-down has not yet been measured, hence, there is no constraint on the powering mechanism and the strength of the magnetic field.  \cite{pelisoli+23} described three kinds of pulses (flares): a short pulse dominating in radio and optical (FWHM$<$10\,s) and a broader pulse seen at optical and X-ray wavelengths, both occurring with the same period, along with a flare with length $\sim$1/4 of a spin cycle that occurred only once at an interpulse phase.

Here, we describe the full body of \srgero observations of \targ obtained between 2020 and 2021 and we present an analysis of a pointed observation with \xmmn obtained in September 2022. Fortunately, it was possible to arrange for simultaneous observations with ULTRACAM at the NTT, which revealed some important insight about the timing properties of the source. Throughout this paper, we use the ephemeris BMJD(TDB) $=59846.050092 + 0.003696715 \times E$ to align our data in phase. This ephemeris puts the broad X-ray pulse at phase zero at the epoch of our observations. It corresponds to phase 0.715 of the spin ephemeris derived by \cite{pelisoli+23}.

\section{\ero\ observations}
\subsection{\ero\ discovery and photometric properties}
Observations by \ero\ on board SRG \citep{predehl+21,sunyaev+21} of the field of \targ\ were carried out four times during the all-sky surveys eRASS1 through  eRASS4  in 2020 and 2021. The survey data were reduced with the \ero\ software eSASS version {\tt eSASSusers\_211214} \citep{brunner+22}, using the latest calibration files. \targ\ was detected as an X-ray source in each of the surveys individually. The data obtained during eRASS1 were used to search for accreting compact white dwarf binaries. A list of 161 high-confidence candidates with matching \gai-sources was compiled using NWAY, a probabilistic cross-matching tool specially trained to identify cataclysmic binaries and related objects \citep{salvato+18,salvato+22}. The location of all the candidates in an optical colour-magnitude and an X-ray to optical colour-colour diagram is shown in Fig.~\ref{f:diag}. Included among them is \targ (\gai designation Gaia DR3 6712706405280342784), shown as a red symbol. As many other cataclysmic binaries (CBs) and candidates, the new WD pulsar is located between the ZAMS and the WD sequence in the CMD. Compared to the other 160 candidates, \targ is relatively red and has a lower X-ray to optical flux ratio, but it is still a factor of 100 above coronal emitters with the same optical colour \citep[we may also make a comparison with Figs. 15 and 16 in ][]{schwope+22}. An identification spectrum showing H$\alpha$ in emission was obtained with the SAAO 1m telescope in July 2022. The much larger body of optical data with higher S/N is described in \cite{pelisoli+23}.

\begin{figure}[th]
\resizebox{\hsize}{!}{\includegraphics{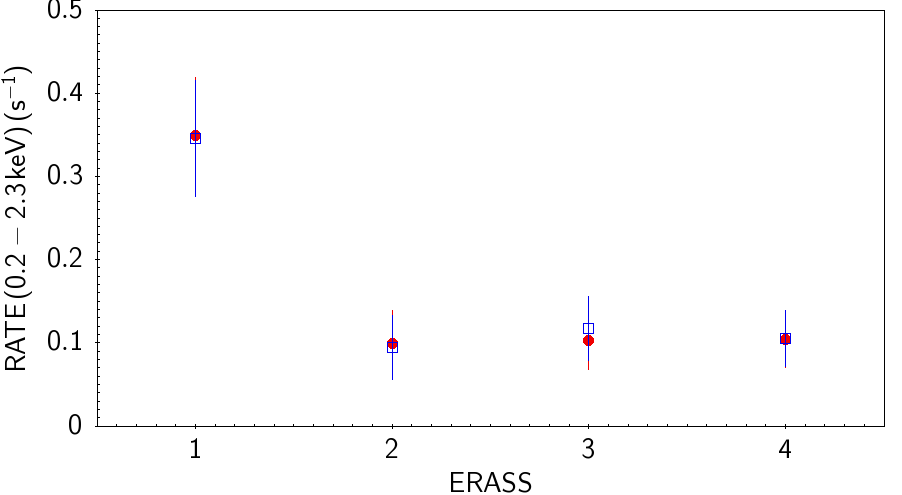}}
\resizebox{\hsize}{!}{\includegraphics{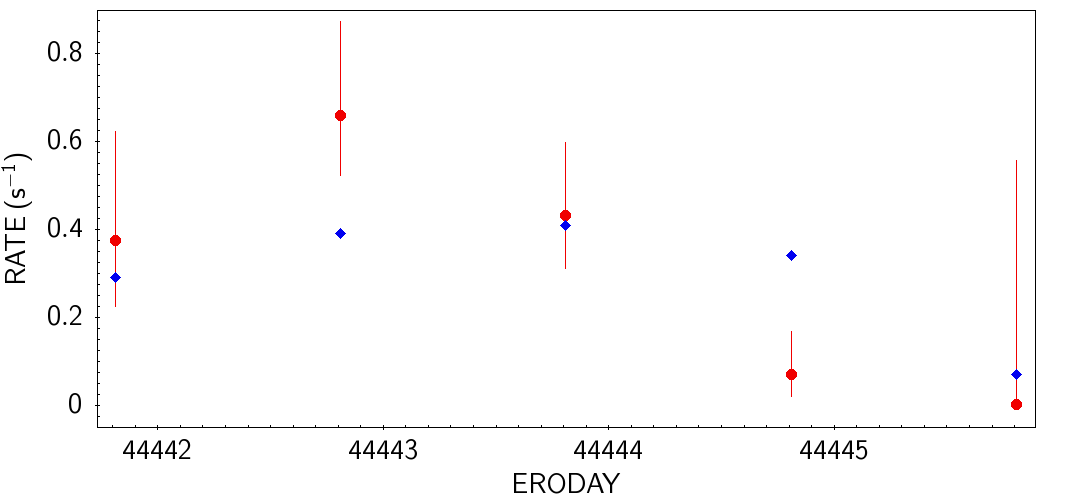}}
\caption{
X-ray variability of \targ observed with \ero. {\it Upper panel:} Long-term X-ray light curve displaying the mean rate per eRASS. Shown are mean count rates from stacked source detection followed by forced photometry to derive fluxes per eRASS (red symbols) and from source detection of the individual surveys (blue symbols). {\it Lower panel:} Mean count rate (red symbols) and exposure time (factor $1/100$, blue symbols) per {\it eRODay} in eRASS1.
}
\label{f:erolc}
\end{figure}

A joint detection on the stacked data from all four eRASS surveys (labeled eRASS:4) revealed the X-ray source at position RA2000 $= 288.057345\degr$ and  DEC2000$=-44.17898\degr$ with a statistical uncertainty of 1.8 arcsec. The X-ray position coincides within 1.2 arcsec with the position of the optical counterpart. The summed vignetting corrected exposure was 370.4\,s (about 80s, 75s, 103s, and 117s in eRASS1, 2, 3,  and 4, respectively) and revealed a total of 57 registered photons. The net exposure before correction was 740 s. The mean eRASS:4 rate was $0.155\pm 0.022 \mbox{s}^{-1}$ in the energy band $0.2-2.3$\,keV that was used for the detection of the X-ray sources. The mean rate corresponds to a flux of $F_{\rm X} \mbox{(0.2-2.3 keV)}= 1.44 \times 10^{-13}$\,\fergs, if the standard ECF (energy conversion factor) is used. The mean rate was higher during eRASS1, $0.35 \pm 0.07$\,\rat, compared to the following three scans where the rate was consistently lower at a rate of $0.10\pm 0.03$\,\rat. The fluxes in the 0.2-2.3 keV band were $3.2, 0.9, 1.1,   1.0 \times 10^{-13}$\,\fergs for eRASS1, 2, 3, and 4, respectively. \ero variability is significant on long and short scales, namely, between different \ero surveys separated by half a year and, in eRASS1,  between {\it eRODays}, as shown in Fig. \ref{f:erolc}. An {\it eRODay} corresponds to one full revolution of the spacecraft (14400\,s), while the length of a scan over a given position on sky is $<$ 40\,s.  An {\it eRODay} is close to but not identical to the orbital period of \targ.

\begin{figure}[t]
\resizebox{\hsize}{!}{\includegraphics{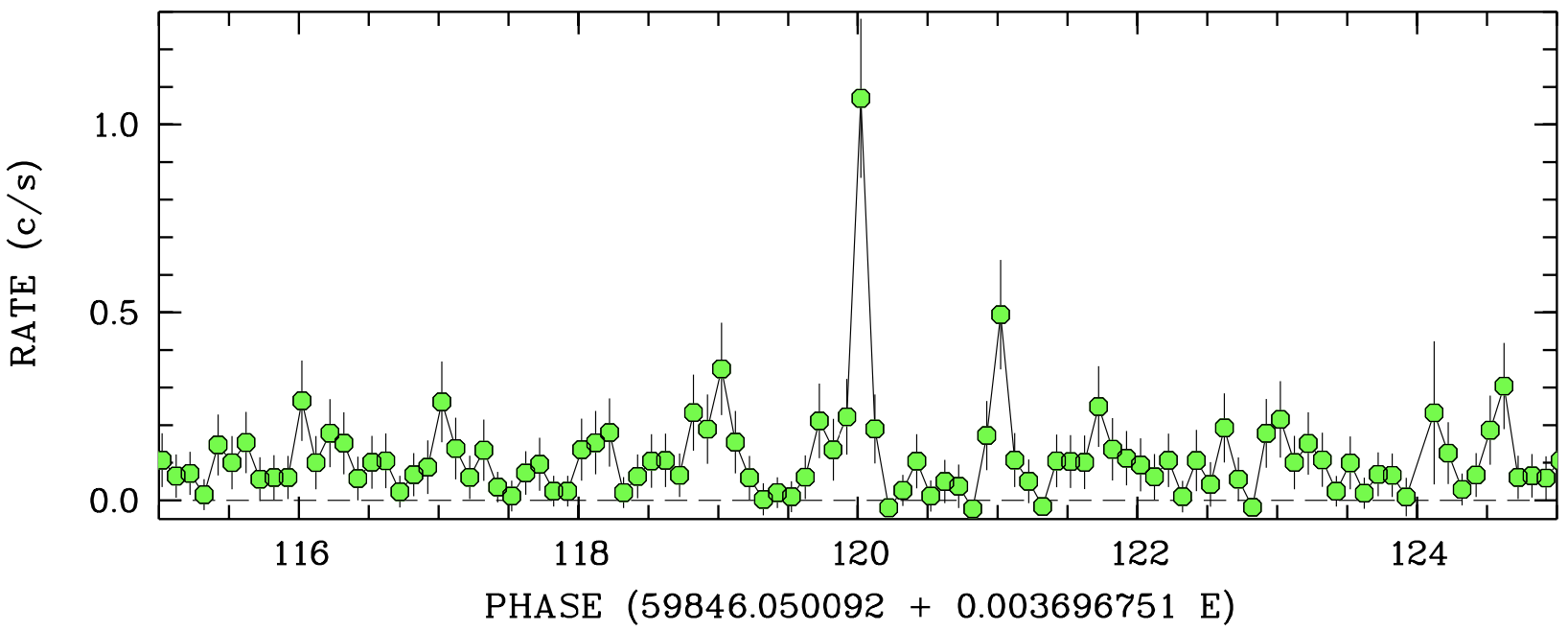}}
\caption{Cutout of the sky-subtracted EPIC-pn light curve (orbital phase $333.226 - 333.444$). Time bins are 32 seconds.}
\label{f:plc}
\end{figure}

\subsection{\ero spectral analysis}
During eRASS1 26 source photons were registered, whereas the 2nd, 3rd, and 4th surveys revealed 7, 12, and 12 photons, respectively. We attempted a rough spectral analysis for eRASS1. Source and background spectra together with the auxiliary files, the RMF and the ARF (redistribution matrix and effective area file) were extracted from the raw photon event file using standard \ero \ procedures. Then, \xspec\ \citep[version 12.12.0,][]{arnaud+96} was used to model the observed spectrum, which was binned with one count per spectral bin. Two spectral models were tested, a thermal plasma emission model \cite[Astrophysical Plasma Emission Code, i.e., the \texttt{apec} model in \xspec terms, ][]{smith+01} and a power-law model, each modified by some amount of cold interstellar matter (\texttt{TBABS}). Both have the capability to reflect the observed spectrum equally well. Given the low number counts, spectral parameters for either model are not well constrained. For a thermal plasma, the lower limit temperature is 1.7\,keV at 90 percent confidence, whereas for a power law, the spectral index lies between 0.6 and 3.0, and the interstellar column density is smaller than $N_{\rm H} < 1 \times 10^{21}$\,cm$^{-2}$.

\begin{figure}[t]
\resizebox{\hsize}{!}{\includegraphics{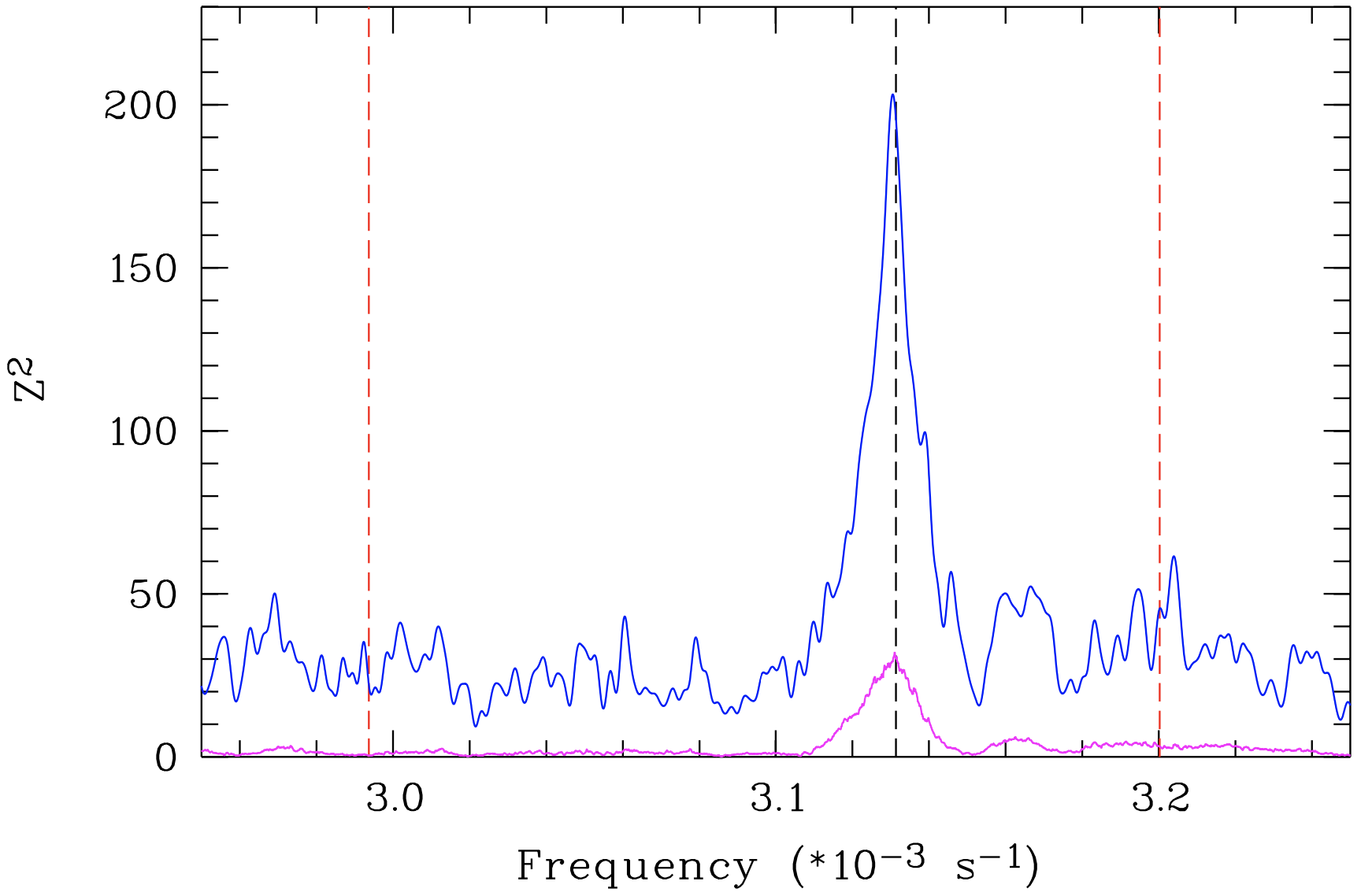}}
\caption{Period search using Buccheri's method (blue curve) and using AOV (magenta). Dashed lines indicate the optically determined spin frequency $\omega_{\rm spin}$ (black), the beat frequency between spin and orbit $\omega + \Omega$ and the sideband frequency $\omega - 2\Omega$.
\label{f:z2m}
}
\end{figure}
\begin{figure}
\resizebox{\hsize}{!}{\includegraphics[]{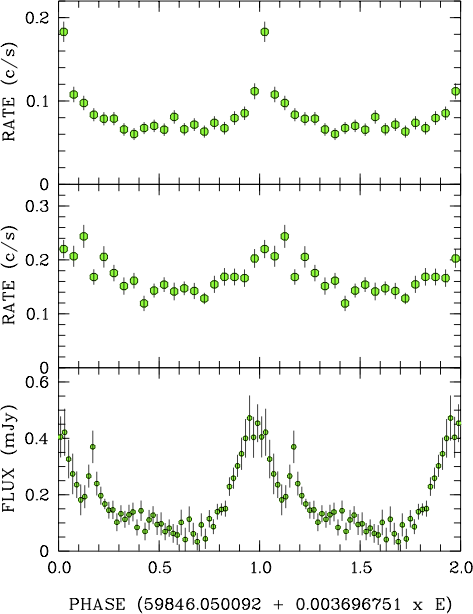}}
\caption{EPIC-pn X-ray, OM/UVM2, and ULTRACAM g-band light curves folded over the spin phase.}
\label{f:xlcs}
\end{figure}

\section{\xmmn observations}
Following our DDT request, \xmmn observed the field of \targ on Sep 17/18, 2022 for a total of 43\,ks. The EPIC X-ray cameras were operated in full-frame mode and the optical monitor was used in fast-imaging mode (time resolution 0.5 s), with the UVM2 filter (effective wavelength 231 nm). The original data were reduced with the latest version of the \xmmn SAS (SAS 20.0) using the most recent calibration files. For the spectral analysis, the data were filtered for intervals of high background rates; the data loss was about $25$\% for EPIC-pn and almost negligible for MOS1 and MOS2. Times of arrival of individual photons were put on the TT time system and corrected to the solar system barycentre using SAS task \texttt{barycen}. Concentric annuli with radii of 30 and 60 arcsec were used to extract source and background photons and to correct the signal in the source region for background contamination. A total of 3430 photons from \targ were collected with EPIC-pn, which gives a mean rate of 0.0825 \rat ($0.2 - 10$\,keV). Fortunately, about 50 min of simultaneous observations \xmmn and  ULTRACAM at the NTT could be ascertained.

The data obtained with the OM were reduced with the SAS task \textit{omfchain}, which generates background-subtracted light curves. Also the OM data were corrected to the solar system barycentre using the \textit{barycen} task. Experiments were made using different time bins to create a light curve. Eventually bins of length 3.19390\,s (corresponding to 100 phase bins of the measured periodicity) were used to then generate a phase-folded light curve with 20 phase bins (shown in Fig.~\ref{f:xlcs}). A fast mode image was created and inspected showing the source at the edge of the fast mode window. This may result in a loss of source photons of up to 50\%.

\begin{figure}[t]
\resizebox{\hsize}{!}{\includegraphics[angle=0]{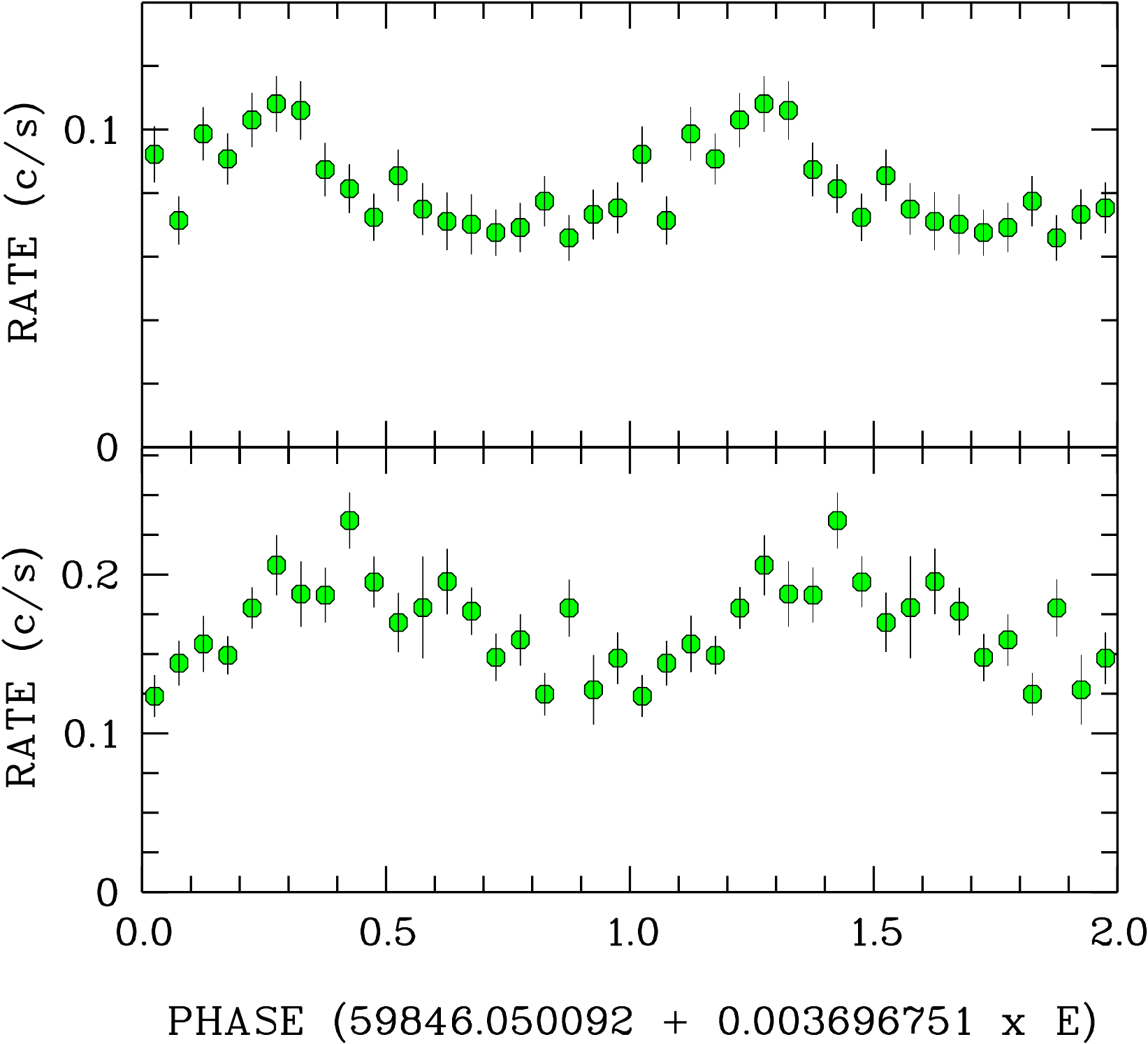}}
\caption{EPIC-pn X-ray and OM/UVM2 light curves folded over the orbital binary period.}
\label{f:xuorb}
\end{figure}

\subsection{X-ray and UV variability}
A cutout of about 50 min of the sky-subtracted, vignetting, and dead time-corrected X-ray light curve is shown in Fig.~\ref{f:plc} (here and in the following, we always use all photons in the 0.2 - 10 keV range). It shows apparent quasi-periodic variability with individual pulses separated by about 5 min. The chosen cutout for the figure contains the most prominent pulse during the whole observation, reaching a peak rate of about 1.07 \rat in a 32\,s time bin (1.69 \rat in a corresponding bin of 16 s in length). 

We searched for periodic variations of the X-ray signal in EPIC-pn using two independent methods. Photons selected in a 30 arcsec region around the source were analysed using Buccheri's $Z_n^2$ statistic \citep{buccheri+83}. The period range investigated was between 100\,s and 450\,s and the resulting periodogram showed one prominent peak at 0.0031305 d$^{-1}$ corresponding to 319.43\,s (Fig.~\ref{f:z2m}). We also used the analysis of variance (AOV) implemented in MIDAS to search for periodic variability. The input used was the sky-subtracted, vignetting and dead time-corrected light curve with 1\,s time bins. The AOV-periodogram, shown also in Fig.~\ref{f:z2m} (magenta line) has its maximum at a period of 319.35\,s, the center of the distribution at 319.52\,s.

\begin{figure}[t]
\resizebox{\hsize}{!}{\includegraphics[angle=00]{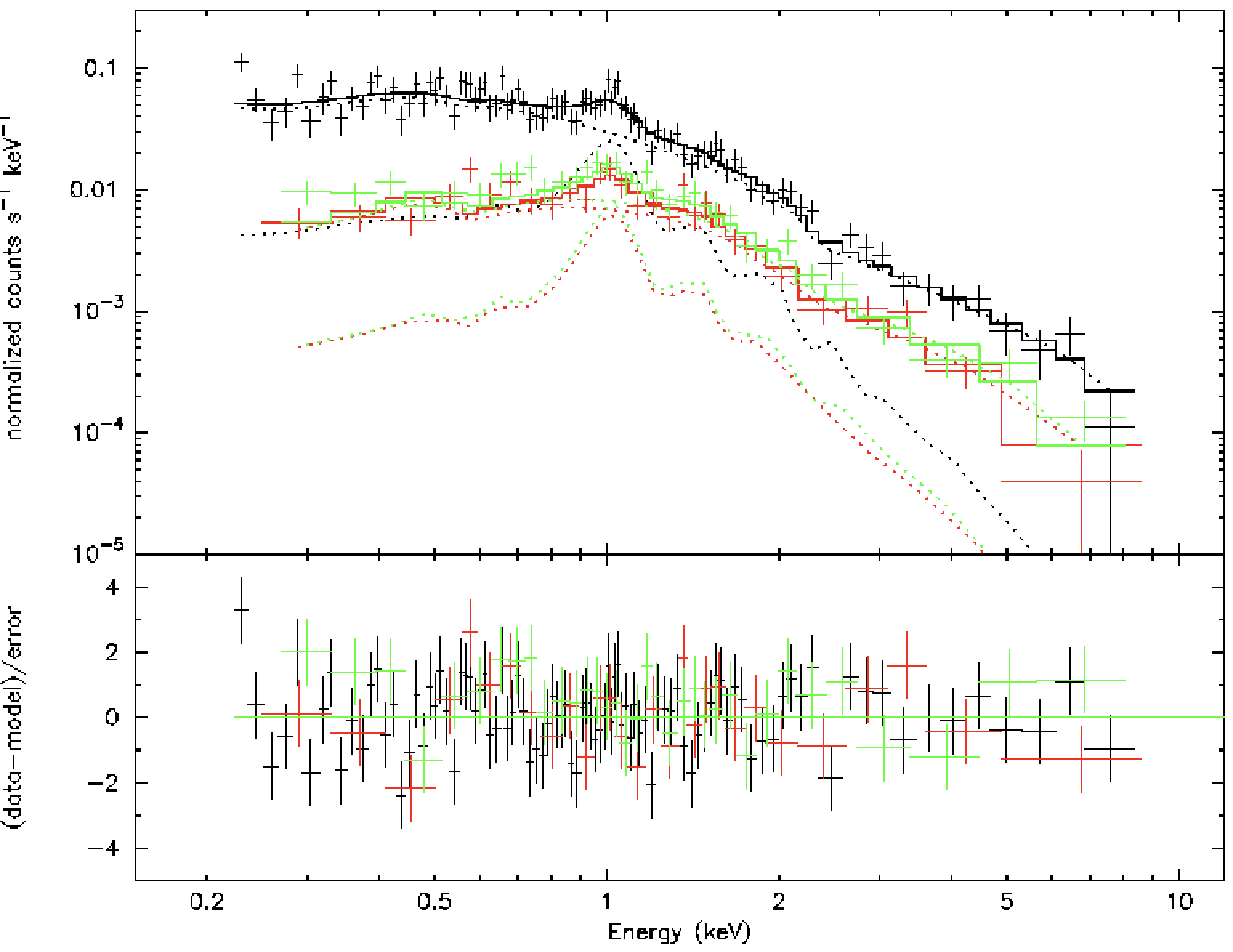}}
\caption{\xmmn spectra obtained with EPIC-pn, -MOS1, and -MOS2 with a power-law plus thermal model fit. Lower panel shows the error-normalized residuals (black: pn, red: MOS1, green: MOS2).}
\label{f:xmmspec}
\end{figure}

Spin phase-folded X-ray, ultraviolet, and optical light curves (ULTRACAM g-band shown, the light curves of the other bands, $r$ and $u$, are almost carbon copies of the $g$-band data) with 20 bins per cycle are shown in Fig.~\ref{f:xlcs}. X-ray and OM data show the same pattern, the broad pulse that appears superposed on a basal flux at about half the spectral flux. The ULTRACAM data additionally show the narrow pulse  occurring a little less than 0.2 phase units after the main pulse. It is not detected  at X-ray wavelengths. Whether the OM data show this feature is difficult to judge given the low count rate. The ephemeris used for folding, MBJD(TDB) $= 59846.050092 + 0.003696751 \times E,$ locates the broad pulse in all wavebands at around $\phi_{\rm spin} = 0$ and, thus, suggest a common physical origin. A smooth interpolated continuum connecting the interpulse minima was subtracted from the ULTRACAM data before phase folding. The mean (basal) rate of the OM in the phase interval 0.3 - 0.7 (spin phase) was $0.152 \pm 0.007$\,\rat, which corresponds\footnote{the conversion factor between counts and measured flux in the UVM2-band is $2.2 \times 10^{-15}$\,\ecf, source \url{https://www.cosmos.esa.int/web/xmm-newton/sas-watchout-uvflux}} to $3.34 \times 10^{-16}$\,\fcgs. 

We searched for orbital phase-dependent variability in the X-ray and UV data from \xmmn by inspecting phase-folded light curves shown in Fig.~\ref{f:xuorb} that were generated using the orbital ephemeris of \cite{pelisoli+23}. The phase-averaged X-ray light curve varies roughly sinusoidally with a maximum at around $\phi_{\rm orb} \sim 0.25$ and has a pulsed fraction $p_f = (f_{\rm max} - f_{\rm min}) / (f_{\rm max} + f_{\rm min})$ of 28\%. The UV light curve is similar in shape, but the bright phase hump a little more extended ($\sim$60\% of the cycle) and centered at a later phase, $\sim$0.45. The UV light curve has the same shape and phasing as the phase-folded optical light curve from \tes \citep[][ their Fig.~1]{pelisoli+23}.

We examined whether the X-ray pulses show a dependency on the orbital phase. We divided the dats set into five orbital phase intervals of same length and searched for the spin pulses in such reduced data sets individually. The pulsations are found at all phases with a tendency to show a larger pulsed fraction between orbital phases 0.0 and 0.4.

\subsection{Spectral analysis}
Mean X-ray spectra for the whole \xmmn observation of \targ were extracted for the three EPIC cameras individually and then fitted jointly. The spectra were binned with a minimum of 25 counts per bin, so that a $\chi^2$ minimization technique could be used to find the best model and to derive the parameter values. Initially a one-component model was tested, but neither a pure thermal emission (\texttt{APEC}) nor a single power law model yielded a satisfactory fit. The fit that reflects the data satisfactorily is a combination of the two emission models modified by a column of cold interstellar matter \texttt{TBABS $\ast$ (APEC + POWL)}. The power law carries most of the flux (almost 90\%), while the thermal model yields the lines around 1 keV to account for spectral residuals left by the power law component alone. The accepted best-fit parameters are listed Table~\ref{t:xfit}, while the data, the model fit (with individual components shown), and the residuals are displayed in Fig.~\ref{f:xmmspec}. The model chosen is the most simple, namely,~the one with least amount of parameters to be determined. The prototype \arsc was fitted with the superposition of several thermal spectra \citep{takata+18,takata+21}. This would be possible for \targ as well. A fit with just two thermal components is not sufficient ($\chi^2_\nu = 1.55$) but the sum of three thermal components with temperatures according to 0.2\,keV, 1.1\,keV, and 4\,keV gives a satisfactory fit ($\chi^2_\nu = 1.07$ for 154 d.o.f.). 

\cite{pelisoli+23} (see their Fig.~3) compiled the spectral energy distribution (SED) from the radio to the X-ray regime using the X-ray spectral model derived here. Integrating this and using the \gai distance, they find the bolometric luminosity to be $\sim$10$^{33}$\,\lum. The X-ray luminosity of \targ, namely, $L_{x} \sim 1.4 \times 10^{30}$\,\lum, is a factor of 3 lower than that of \arsc, while the fraction $L_{\rm X} / L_{\rm bol}$ is only $\sim$0.14\% for \targ. However, this fraction seems to be slightly higher (on the order of 1-2\%) for \arsc \citep{takata+21}.

\begin{table}
\caption{Parameters of best-fit models to the mean EPIC spectra. The $\kappa$ is the power law index. 
\label{t:xfit}}
\begin{tabular}{lc}
\hline       
\hline                     
\multicolumn{2}{l}{Model: \texttt{TBABS*(APEC + POWL)}}           \\
\hline
Parameters               &                                \\
\hline   
$N_{\rm H}$($10^{20}$cm$^{2}$)  & $5 \pm 2$   \\
$kT_{\rm APEC}$(keV)            & $1.24^{+0.11}_{-0.10}$     \\
Norm$_{\rm APEC}$ & $(1.5 \pm 0.5) \times 10^{-5}$ \\
$\kappa_{\rm PL}$ & $2.14\pm0.11$ \\
Norm$_{\rm PL}$ & $(3.11 \pm 0.38) \times 10^{-5}$ \\
$\chi_\nu^{2}$ \,\,($\chi^{2}$ / d.o.f)            & 1.12(173/154)             \\
\hline
\multicolumn{2}{l}{Observed Fluxes $(10^{-13}\fergs)$  }                 \\
$F_{0.2-2.3} $  & $1.05 \pm 0.04$   \\
$F_{0.2-10}  $  & $1.66 \pm 0.09$        \\
\hline
\multicolumn{2}{l}{Unabsorbed Fluxes, best-fit parameters $(10^{-13}\fergs)$}\\
$F_{0.2-2.3} $  & $1.5$   \\
$F_{0.2-10}  $  & $2.1$        \\
$F_{\rm bol} $  & $12.3$ \\
\hline
$L_{x}$(0.2-10 keV, $(D/237\,\rm{pc})^2 10^{30}$ erg s$^{-1}$) & $1.4$      \\
\hline
\end{tabular}
\end{table}

\section{Results and discussion}
We analysed the currently available X-ray data of \targ from \ero and \xmmn. First, it is worth noting that \targ was not detected by \ros, \swi, or by the \xmmn slew survey. \targ shows X-ray variability on various time scales, the longest investigated being over half a year, namely, the time interval between individual eRASS scans. The mean brightness during eRASS1 with a flux of $3.3 \times 10^{-13}$\,\fergs dropped to $\sim$$1 \times 10^{-13}$\,\fergs at all subsequent occasions. During eRASS1, \targ was also variable on the {\it eRODay} scale (which roughly correspond to the orbital period). Pointed \xmmn observations revealed periodic variability of the X-ray and the UV signals on the spin period of 319.4\,s. Phase-folded light curves on the spin period show a basal flux and the broad pulses seen also in the optical.  Their maximum flux is a factor $\sim$2 higher than the basal flux. Phase-folded light curves on the orbital period have a pulsed fraction of 28\% and a bright phase lasting about 0.5 phase units, and that is roughly centered on phase 0.25. The narrow pulse, which is prominent at radio and optical wavelengths was not discovered in the 130 spin cycles covered by \xmmn.  A broad interpulse flare, which could be an accretion event, seen in the optical and X-ray, would perhaps remained unnoticed without its prominent appearance in the X-ray data.

\cite{pelisoli+23} explained the period of 319\,s as the spin period of the white dwarf.  The predicted beat period is 312.5 \,s, while (almost) no signal is seen in the periodograms at this period, neither in this work nor in \cite{pelisoli+23}. This is unexpected, because the emission lines, which are undoubtedly originating from the irradiated side of the donor star,  show clear flux modulations that hint to a pronounced irradiation effect of the secondary star. These are expected to occur on the beat period.  An observational confirmation is outstanding because the time coverage of the spectroscopic data was insufficient to distinguish between the beat and spin periods. The X-ray data show clear evidence for variability on the suspected spin period whereas there is at most marginal evidence for the beat period in the X-ray periodogram (Fig.~\ref{f:z2m}).

A couple of questions remain. Narrow optical pulses during the joint ULTRACAM/\xmmn campaign are thought to be of synchrotron origin from the WD, but they have no X-ray counterpart. The narrow pulses are the only signal seen at radio wavelengths. The optical pulses were observed in cycle 331 at orbital phases 0.0526, 0.746, 0.967, and 0.1186 (with increasing flux) but were not observed at later phases, although the observations were continued through further six spin cycles up to orbital phase 0.26. Some parameters of the narrow flares are puzzling, in particular with regard to what is causing their phase dependence and what their genuine energy distribution is.

If the pulsed X-ray emission is due to synchrotron beaming, we must consider what the source of the basal flux is, for instance, whether it may be due to weak accretion. The measured X-ray luminosity is low, much lower than what a Roche lobe-filling cataclysmic variable at this orbital period would typically exhibit. An active secondary star might be the source of a wind that is captured by the strong magnetic field of the WD, as in the LARPs/PREPs \citep{schwope+02, schwope+09}. This may lead to accretion-induced X-ray emission as in other LARPs \citep[e.g.][]{vogel+07}, although the accreted matter is guided to the footpoints of the magnetic fields and such an accretion geometry typically leads to pulsed X-ray emission. An argument in favour of a scenario that involves accretion may be derived from the observed optical/X-ray flare. The basal flux might be composed by a number of unresolved miniflares. Flares or flare-like accretion might also be the cause of the observed variability between {\it eRODays }during eRASS1.

Orbital phase-dependent variability is observed at X-ray and UV wavelengths. It is important to take care to avoid overinterpreting the data, as they cover just three orbital cycles and phase-dependent behavior might be observed where it is not actually present. However, it appears as if the UV would be bright around phase 0.5. We consider whether this be the result of the irradiation of the donor star, which would then radiate at a level that corresponds to the flux difference between the mean flux at phases 0.0 and 0.5 (i.e.~$\sim$$1.7 \times 10^{-16}$\,\fcgs). The X-ray light curve peaks earlier in orbital phase at around phase 0.25. It seems to display a slow rise and a faster decline from the maximum. The light curve asymmetry is reversed from left to right in comparison to \arsc. 

There are other dissimilarities that ought to be considered when making comparisons with \arsc. In particular, for \arsc, the X-ray and UV light curves appear similar, especially with regard to the phase of the minimum flux, which is the same. The curves in \targ appear dissimilar. The spin phase-folded light curves in \arsc are double-peaked, while those in \targ are single-peaked. \arsc seems to have a strong thermal component in its X-ray spectrum, whereas that of \targ could be of non-thermal origin. The question of whether those differences can be explained simply by different geometries or whether they require different physical arguments can be addressed once data with higher signal-to-noise ratios are available.

\begin{acknowledgements}
We thankfully acknowledge constructive criticism of an anonymous referee.

Support of the Deutsche Forschungsgemeinschaft under project Schw536/37-1 is gratefully acknowledged.

This work is based on data from eROSITA, the soft X-ray instrument aboard SRG, a joint Russian-German science mission supported by the Russian Space Agency (Roskosmos), in the interests of the Russian Academy of Sciences represented by its Space Research Institute (IKI), and the Deutsches Zentrum für Luft- und Raumfahrt (DLR). The SRG spacecraft was built by Lavochkin Association (NPOL) and its subcontractors, and is operated by NPOL with support from the Max Planck Institute for Extraterrestrial Physics (MPE).

The development and construction of the eROSITA X-ray instrument was led by MPE, with contributions from the Dr. Karl Remeis Observatory Bamberg \& ECAP (FAU Erlangen-Nuernberg), the University of Hamburg Observatory, the Leibniz Institute for Astrophysics Potsdam (AIP), and the Institute for Astronomy and Astrophysics of the University of Tübingen, with the support of DLR and the Max Planck Society. The Argelander Institute for Astronomy of the University of Bonn and the Ludwig Maximilians Universität Munich also participated in the science preparation for eROSITA.

The eROSITA data shown here were processed using the eSASS/NRTA software system developed by the German eROSITA consortium.
\end{acknowledgements}

\bibliographystyle{aa}
\bibliography{j1912_x}

\end{document}